\documentclass[lettersize,journal]{IEEEtran}
\usepackage{amsmath,amsfonts}
\usepackage{algorithmic}
\usepackage{algorithm}
\usepackage{array}
\usepackage[caption=false,font=normalsize,labelfont=sf,textfont=sf]{subfig}
\usepackage{textcomp}
\usepackage{stfloats}
\usepackage{url}
\usepackage{xcolor}
\usepackage{lettrine}
\usepackage{verbatim}
\usepackage{graphicx}
\usepackage{cite}
\usepackage[whole]{bxcjkjatype}
\usepackage{makecell} 
\begin{document}

\title{A 55-nm SRAM Chip Scanning Errors Every \\125 ns for Event-Wise Soft Error Measurement}

\author{
Yuibi Gomi,~\IEEEmembership{Student Member,~IEEE,}
Akira Sato,
Waleed Madany,~\IEEEmembership{Member,~IEEE,}
Kenichi Okada,~\IEEEmembership{Fellow,~IEEE,}
Satoshi Adachi,
Masatoshi Itoh,
Masanori Hashimoto,~\IEEEmembership{Senior Member,~IEEE}
\thanks{
Manuscript received 11 April 2025; revised 21 June 2025.
}
\thanks{This work is supported by the Grant-in-Aid for Scientific Research (S) from Japan Society for the Promotion of Science (JSPS) under Grant 24H00073 and JP19H05664.
}
\thanks{Yuibi Gomi and Masanori Hashimoto are with the Department of Informatics, Kyoto University, Kyoto 606-8501, Japan (email: hashimoto@i.kyoto-u.ac.jp).
}
\thanks{Akira Sato is with the Department of Physics, Osaka University, Toyonaka 560-0043, Japan.}
\thanks{Waleed Madany and Kenichi Okada are with the Department of Electrical and Electronic Engineering, Institute of Science Tokyo, Tokyo 152-8550, Japan.}
\thanks{Satoshi Adachi and Masatoshi Itoh are with the Research Center for Accelerator and Radioisotope Science (RARiS), Tohoku University, Sendai 980-8578, Japan.}
}

\markboth{IEEE SOLID-STATE CIRCUITS LETTERS,~Vol.~XX, No.~X, August~20XX}%
{Shell \MakeLowercase{\textit{et al.}}: A Sample Article Using IEEEtran.cls for IEEE Journals}

\IEEEpubid{0000--0000/00\$00.00~\copyright~2021 IEEE}

\maketitle

\begin{abstract}
We developed a 55 nm CMOS SRAM chip that scans all data every 125 ns and outputs timestamped soft error data via an SPI interface through a FIFO.  The proposed system, consisting of the developed chip and particle detectors, enables event-wise soft error measurement and precise identification of SBUs and MCUs, thus resolving misclassifications such as Pseudo- and Distant MCUs that conventional methods cannot distinguish.
An 80-MeV proton irradiation experiment at RARiS, Tohoku University verified the system operation. Timestamps between the SRAM chip and the particle detectors were successfully synchronized, accounting for PLL disturbances caused by radiation. Event building was achieved by determining a reset offset with sub-ns resolution, and spatial synchronization was maintained within several tens of micrometers.
%
\end{abstract}


\section{Introduction}
\IEEEPARstart{S}{oft} errors from radiation-induced transient faults challenge modern systems, especially in safety-critical applications such as autonomous driving. Although soft errors in a single device are rare, their occurrence becomes inevitable given the large number of devices.

Soft errors are part of single-event effects (SEEs), and they typically include single-event upsets (SEUs) and single-event transients (SETs).  An SEU occurs when radiation strikes an SRAM cell or flip-flop, generating electron–hole pairs that flip a stored bit. Charged particles deposit charge directly, while neutrons deposit charge via the generated secondary ions, potentially causing silent data corruption or detected unrecoverable errors.
SEUs appear as either single bit upsets (SBUs) or multiple-cell upsets (MCUs), with MCUs possibly undermining error correction codes (ECC) \cite{1269335}. In space, protons ionize directly and via secondary ions \cite{659039}
; on Earth, neutrons dominate \cite{10855824}, with muons also contributing via nuclear reactions \cite{5658005, 10891050}.

The conventional irradiation test writes values into SRAMs, irradiates for a fixed period, and then reads out accumulated bit flips. This approach suffers from issues such as \textit{Pseudo MCU} where independent SEUs are aggregated as a single MCU and \textit{Distant MCU} where bit flips from one particle are misidentified as multiple SEUs. 
Seifert et al. mitigated the Pseudo MCU issue by continuously reading SRAMs during irradiation to submicron processes and FinFET structures
\cite{seifert2008multi,seifert2012soft}
, but Distant MCU remains unaddressed.

This paper proposes a system that simultaneously acquires particle hit timing and location data via a detector and captures SRAM bit flip data.  By combining these data streams, our method eliminates Pseudo MCUs and accurately measures SEUs event by event, including Distant MCUs. For this system, we newly designed a 55-nm SRAM chip that scans for errors every 128 clock cycles at a clock frequency of 540 to 1025 MHz and sends error timing and location data to a PC host via FIFO. Proton-beam irradiation experiments verified the chip and system operations.

\IEEEpubidadjcol

\section{An Event-wise Soft Error Measurement System}
\begin{figure}[!tbp]
\centerline{\includegraphics[width=220pt]{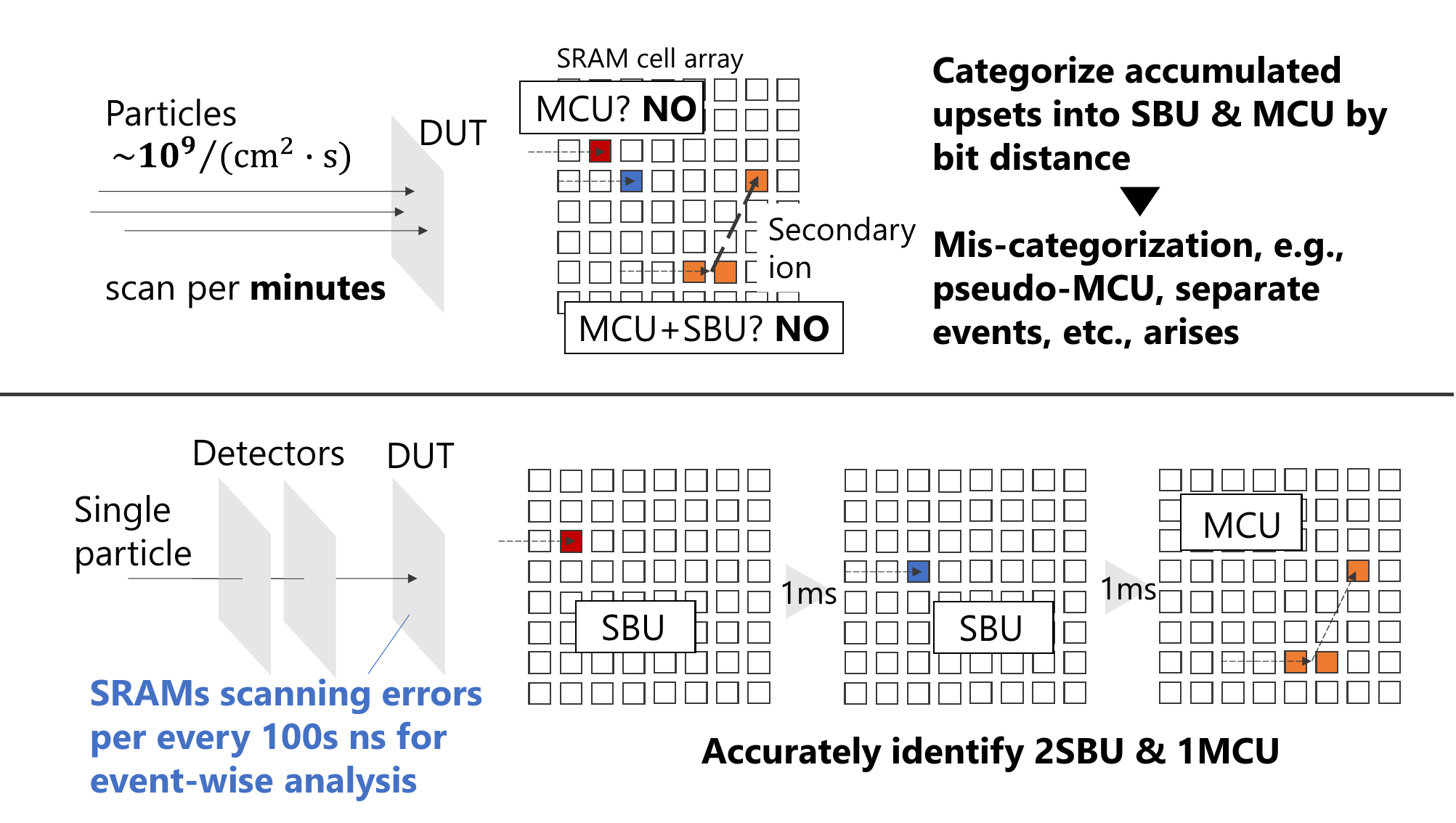}}
\caption{
Conventional (upper) and proposed (lower) methods in radiation testing experiments. The proposed method enables more accurate identification of SBU and MCU events than the traditional upset accumulation.
}
\label{fig:convvsproposed}
\end{figure}

We have developed an event-wise soft error measurement system, as shown in Fig.~\ref{fig:convvsproposed}. Conventional soft error experiments accumulate bit flips over a fixed irradiation period and then map them onto a physical bitmap to classify SEUs into SBUs or MCUs. While efficient, this approach relies solely on location information, leading to the issues described earlier: Pseudo MCU and Distant MCU.
Pseudo MCUs occur when unrelated SEUs appear nearby due to prolonged irradiation.
Distant MCUs result from one particle causing distant bit flips via nuclear reactions.
Both issues arise from using only spatial data without timing.
To address these issues, the proposed system integrates a dedicated high-speed SRAM chip, a plastic scintillator, and a Si detector. When a particle strikes the system, the scintillator and the Si detector detect its time and position while the SRAM chip records bit flip locations and timestamps. These four data samples are collectively tracked to identify the particle responsible for each bit flip and group all bit flips caused by the same particle.
This approach establishes precise temporal and spatial correlations between bit flips and radiation events, eliminating misclassifications such as Pseudo and Distant MCUs. The key challenges are designing the dedicated SRAM chip and ensuring detector–DUT time synchronization, which are detailed in the following subsections.

\subsection{Chip Design}
\label{subsec:chip}

A key design specification is the scan interval, which sets the temporal resolution of SEU events based on the SRAM word count and clock frequency. Thus, an SRAM with fewer words and a higher-frequency PLL are preferable.

\begin{figure}[!tbp]
\centerline{\includegraphics[width=280pt]{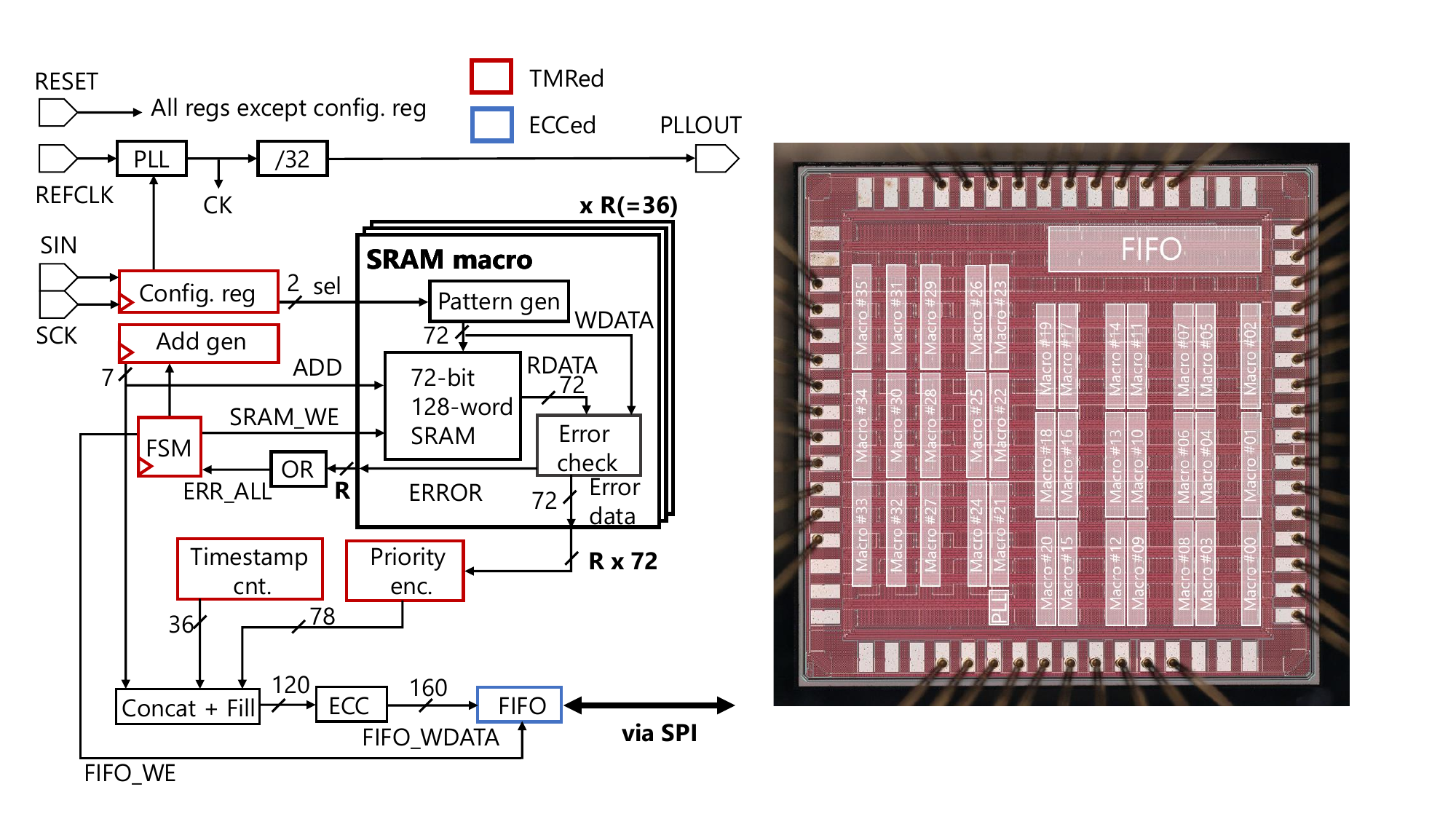}}
\caption{
A block diagram and a photograph of the designed chip.
}
\label{fig:chipdiagramandphoto}
\end{figure}

\begin{figure}[!tbp]
\centerline{\includegraphics[width=150pt]{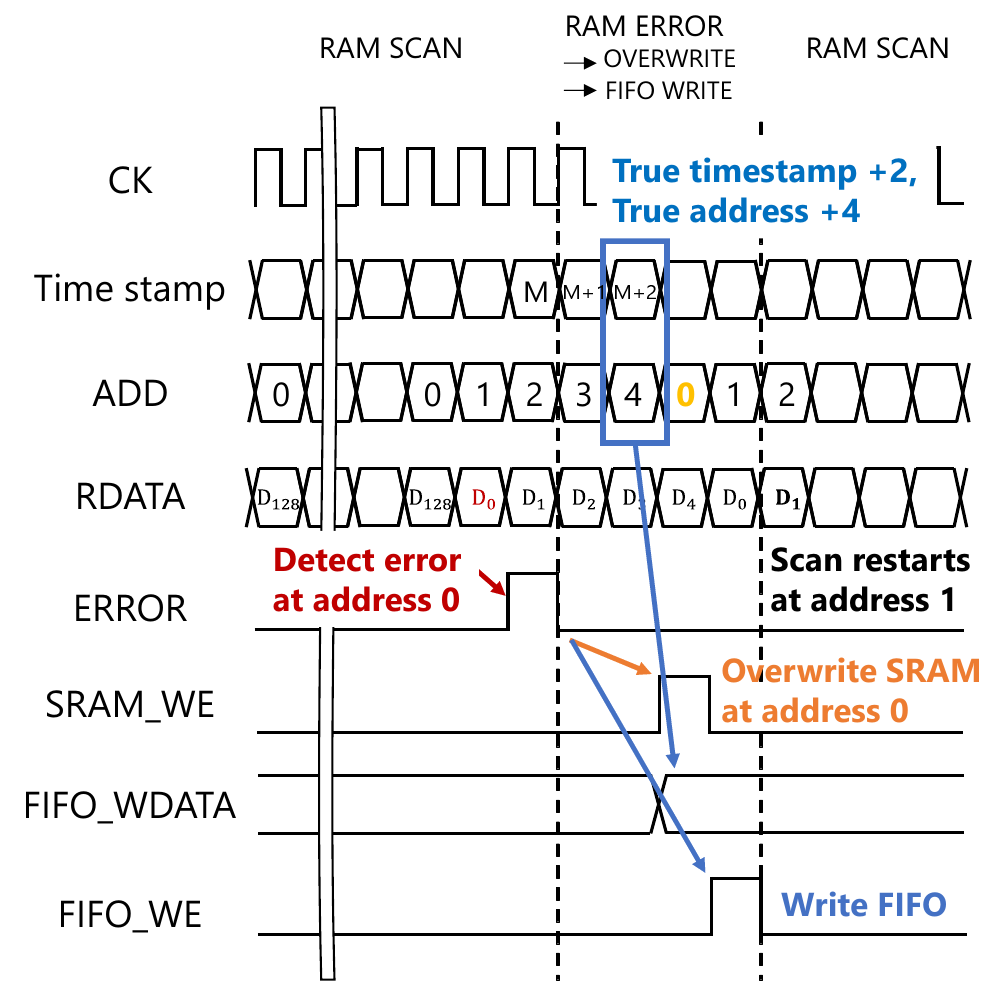}}
\caption{
Timing chart of the designed chip.
}
\label{fig:timingchart}
\end{figure}

Figs.~\ref{fig:chipdiagramandphoto} and \ref{fig:timingchart} show a chip block diagram and timing chart, highlighting the rapid error-checking process and secure transfer of error timing and location to the FIFO. The chip contains 36 SRAM macros, each with a one-port, 72-bit, 128-word SRAM, a data pattern generator, and an error-checking mechanism. The error checker outputs an ERROR signal and a 72-bit error data signal with 1s indicating bit flip locations, while a global ERR\_ALL signal is generated by OR-ing all ERROR signals. Redundancy measures, including triple modular redundancy and ECC, ensure functionality and protect error data. 
A photograph of the 55-nm fabricated chip is shown in Fig.~\ref{fig:chipdiagramandphoto}; the chip measures 3.88 mm$^2$, each macro occupies 0.0363 mm$^2$, and the macros cover 33.7\% of the core area.

When ERR\_ALL is high, the finite state machine (FSM) transfers error data to the FIFO and clears errors to resume scanning. For instance, if an error occurs at address 0 (as shown in Fig.~\ref{fig:timingchart}), FIFO\_WDATA, comprising error data, address, timestamp, and macro number, is assembled using a priority encoder and a 36-bit counter, then stored via FIFO\_WE. The FSM also controls the address generator to overwrite the expected value at address 0, while a priority encoder handles simultaneous errors at the same address in different macros. The FIFO supports asynchronous SPI reading, and if no error is detected, the address increments each cycle, ensuring that all 128 words are checked in 128 cycles.

The PLL, implemented with standard cells as in \cite{9258394}, generates a 540 MHz to 1025 MHz clock at 0.9 V to drive the system. 
Therefore, all cells can be checked within 125 ns at 1025 MHz operation and 237 ns at 540 MHz.
This clock is divided by 32 and output as PLLOUT, which is monitored to verify lock status and assist in timestamp retrieval.
At 125~ns scanning, the additional power overhead is 0.50~W.
A higher clock improves temporal resolution and allows efficient measurement at high beam intensities.
It is also essential when our chip is irradiated by neutrons instead of protons, as external detectors cannot be employed and timing is the only means to distinguish SEUs. A faster clock reduces misclassification.

For maximum particle flux, all SRAM contents must be read before the next particle arrives. At 540 MHz, 
if particles arrive at intervals longer than 237 ns, a beam rate of up to 4.21 MHz per macro is feasible, corresponding to 6.88 $\times10^{10}$ /cm$^2$/s. 
Meanwhile, particle arrivals are inherently random.
This suggests that the beam flux may need to be reduced to avoid hits occurring within short time intervals.
However, even at low flux, particle arrival timing cannot be guaranteed.
By discarding hits with intervals shorter than 237 ns, our system can still operate at high flux while preventing Pseudo MCUs.

\subsection{Timing Synchronization}\label{timingsynchronization}
It is imperative to correlate soft error data timestamps, which are sourced from the SRAM, PLL, and FIFO in the chip, with the detector data. 
To accurately reproduce individual events, system-wide timing synchronization is essential.
Considering that these physical phenomena occur at the nanosecond scale, synchronization must be performed at the same level.
For this, we feed the same reference clock signal to both the detectors and the DUT and initiate their resets using the same reset signal. 

\begin{figure}[!tbp]
\centerline{\includegraphics[width=180pt]{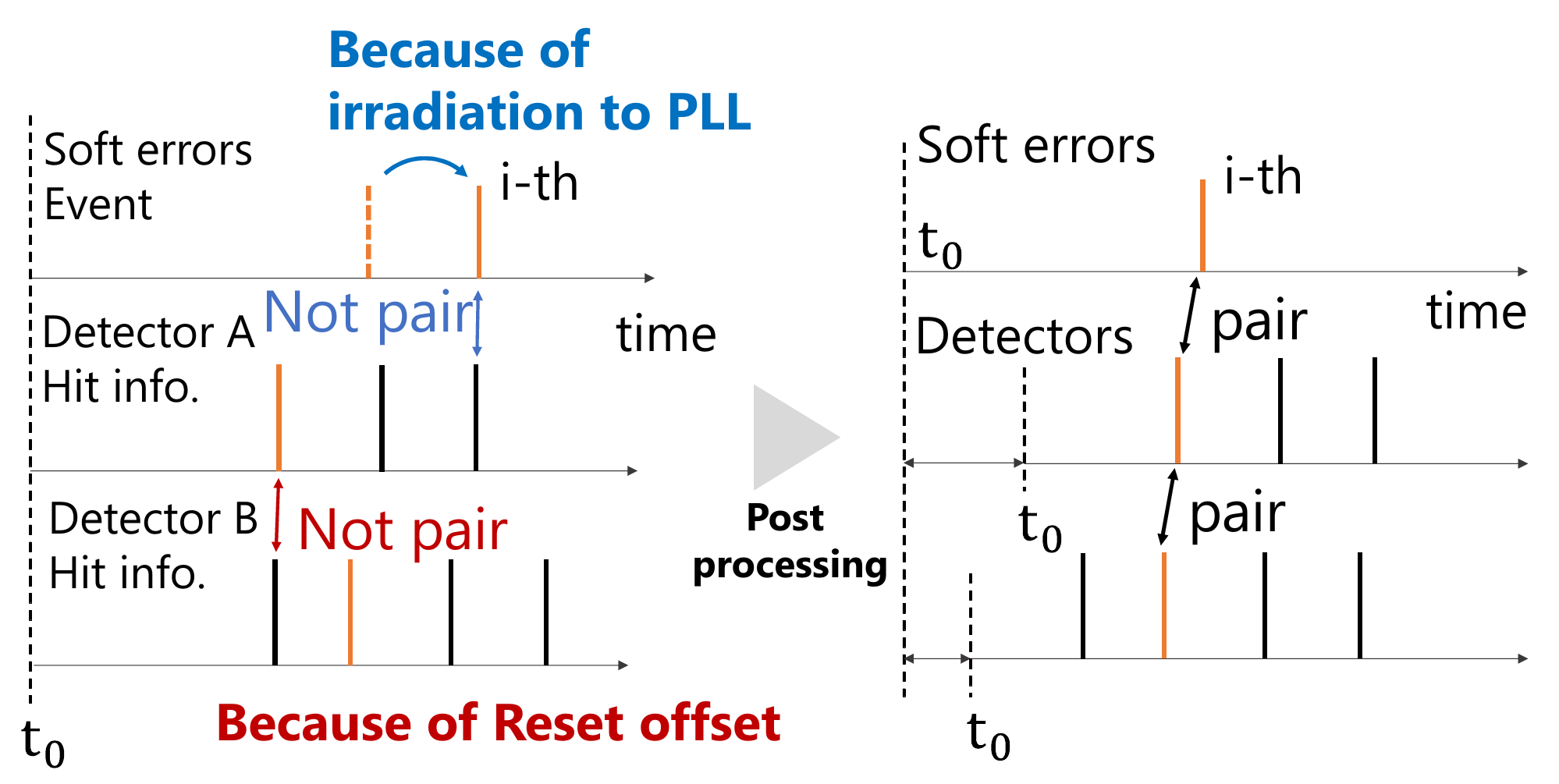}}
\caption{
Pairing of soft errors and hit information from detectors through post-processing. Discrepancies arise due to radiation exposure of the PLL, which generates the internal clock, and the reset offset in the measurement system.
}
\label{fig:postprocessing}
\end{figure}
\begin{figure}[!tbp]
\centerline{\includegraphics[width=140pt]{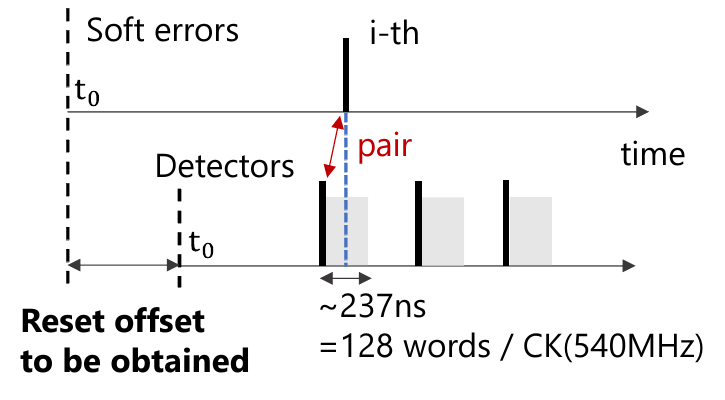}}
\caption{
Synchronization of soft errors and detector hits. The reset offset is determined to pair all upsets and detector hit information. For a 128-word SRAM and a 540 MHz clock, the detection-to-identification time is $\le$237 ns.
}
\label{fig:resetoffsetprocessing}
\end{figure}

Fig.~\ref{fig:postprocessing} highlights two issues in associating soft errors with detector hits.
The first problem is that the timestamps of the soft errors will be out of synchronization with the entire system when the PLL in the DUT is affected by radiation.
For resolving this issue and synchronizing the timing, PLLOUT, an external output signal obtained by dividing the PLL clock by 32, and the 50 MHz REFCLK are tracked by field programmable gate array (FPGA) counters, with values retrieved every 40 ms via serial communication. 
When the PLL is locked, actual time is directly derived from their proportional relationship. 
If unlocked, linear interpolation (LERP) estimates the actual time.
The second problem is that the system-wide reset signal does not reach every device at the same time. This inherent time lag necessitates determining a reset offset during post-processing. 
Fig.~\ref{fig:resetoffsetprocessing} illustrates how the i-th soft error and detector hit are paired within the possible error occurrence-to-detection time of 237~ns, which is the necessary time to scan all bits.
By sweeping the reset offset, we find the offset candidates that enable complete pairing of all events.
Among them,  the one closest to the average time margin is selected.
Here, the average time margin is defined as the expected average discrepancy across all event pairs.
For example, at 540 MHz operation, since the time discrepancy between soft error and detector timestamps can be as large as 237 ns and is uniformly distributed, the expected average time discrepancy is 118.5 ns.

%

\section{Experiments for System Validation}
\subsection{Setup}
\begin{figure}[!tbp]
\centerline{\includegraphics[width=200pt]{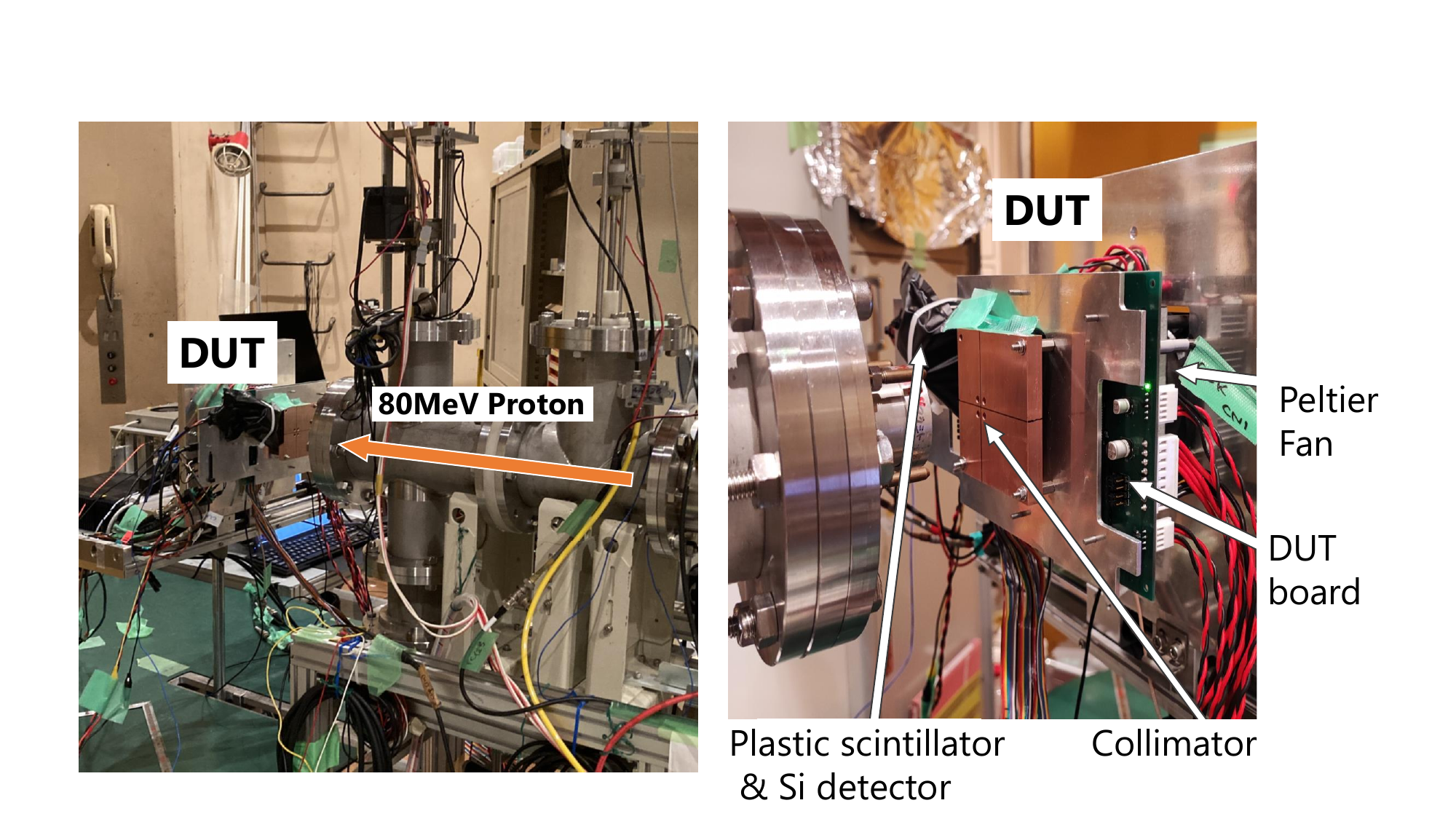}}
\caption{
Irradiation experiment setup at Tohoku University RARiS.
}
\label{fig:experimentalsetup}
\end{figure}

Fig.~\ref{fig:experimentalsetup} presents the experimental setup at the research center for accelerator and radioisotope science (RARiS), Tohoku University. 
An 80-MeV proton beam was used for irradiation to induce nuclear reactions.
The irradiated system includes a collimator with four holes for four chips, a plastic scintillator with 2 ns resolution, a Si detector with 55 $\mathrm{\mu}$m resolution, the DUT board, a Peltier cooler, and a fan. 
Although the board housed 24 chips, only four were activated, with one operating the PLL for analysis. 
Each 55-nm CMOS chip contained 0.332~Mbit SRAM.
Before irradiation, all SRAM cells were initialized to 0, and the supply voltage was set to 0.9~V. 
When the chip operated above 500~MHz, its current draw exceeded 1~A, causing the temperature to rise above 80~°C and potentially affecting PLL locking. To maintain stable operation, a Peltier cooler, which kept the temperature at 20~°C, was attached to the back side of the DUT board.
The detectors and DUT were placed with a 12.8-mm interval.
External modules provide the reset signal and reference clock.

\subsection{Results and Discussion}

\begin{figure}[!tbp]
\centerline{\includegraphics[width=180pt]{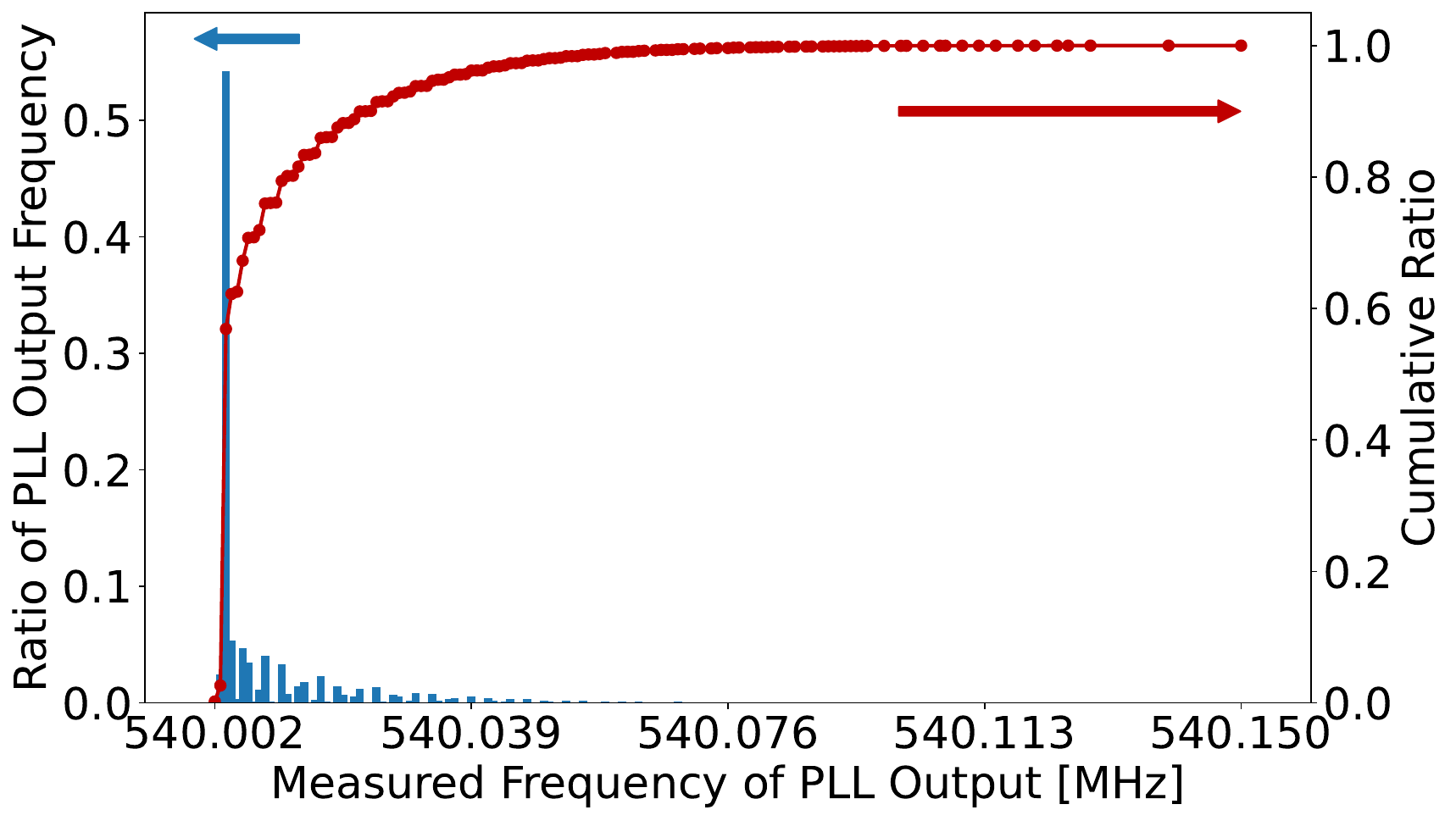}}
\caption{
The occurrence frequency and cumulative ratio of the PLL output observed at the PLLOUT signal. 
The PLL output is obtained by multiplying the frequency of the divided PLLOUT signal.}
\label{fig:plloutput}
\end{figure}

By monitoring the PLLOUT signal with an FPGA, we successfully measured and corrected the timestamps. Fig.~\ref{fig:plloutput} shows the occurrence frequency and cumulative distribution of the measured PLL output frequency. The PLL output frequency was obtained by multiplying the frequency of the divided PLLOUT signal with 32. While 540.004 MHz accounted for approximately 54\% of the measurements, there were instances where the PLL output frequency deviated from this value. The maximum observed frequency was 540.1496 MHz, corresponding to an error of 0.027\%, which might be acceptable for typical operations. 
However, in this experiment, time synchronization at the nanosecond scale is required, making even a 1 Hz deviation significant. To address this, we utilized the post-processing method 
to calculate the actual time and achieve synchronization with the detector's timing. 

\begin{figure}[!tbp]
\centerline{\includegraphics[width=180pt]{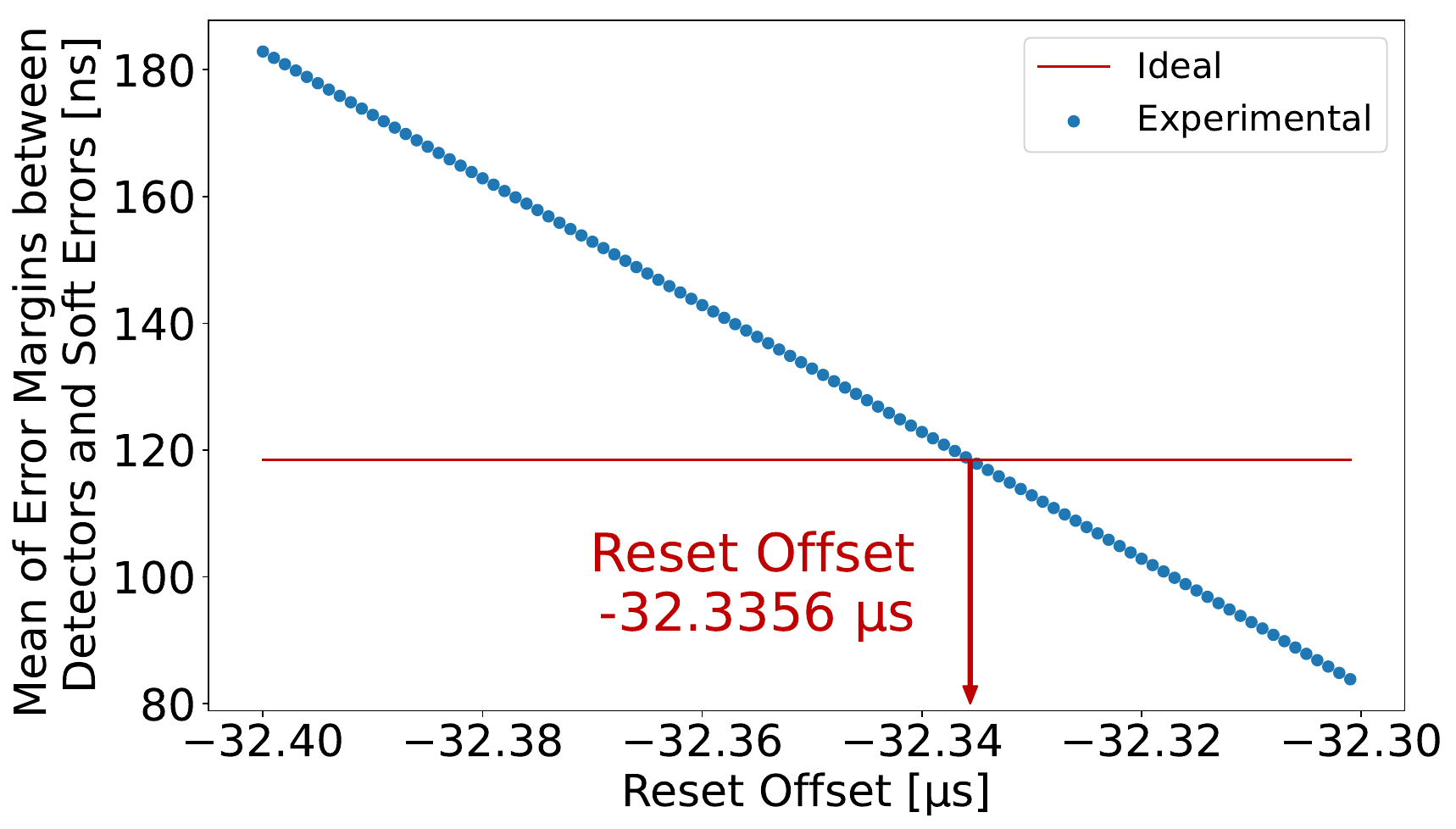}}
\caption{
The mean of the margins between soft errors and detector hits as a function of the reset offset. 
The reset offset at which the mean margin matches the ideal value is defined as the true reset offset in this experiments.
}
\label{fig:resetoffset}
\end{figure}

Subsequently, using the post-processing method
shown in Fig.~\ref{fig:resetoffsetprocessing}
, we determined the reset offset that pairs all soft error events with the detector hit events, where six soft error events were measured in this test run. Initially, the reset offset was varied in 1-ns increments to identify the range of reset offset that successfully paired all soft error events with detector hit events. In the identified range of reset offset, we determined a single reset offset value by calculating the margins between the timestamps of all soft error and detector hit events, and averaging these margins. The results are shown in Fig.~\ref{fig:resetoffset}. In this experiment, the margin is within 237 ns, and the ideal average value for all margins is 118.5 ns, because the time it takes for a particle to pass through the detector and cause an error, and for the error to be detected, is equally probable for all errors. 
Based on this analysis, the reset offset of -32.3556 $\mathrm{\mu}$s was identified as the unique reset offset for this experiment.

\begin{figure}[!tbp]
\centerline{\includegraphics[width=180pt]{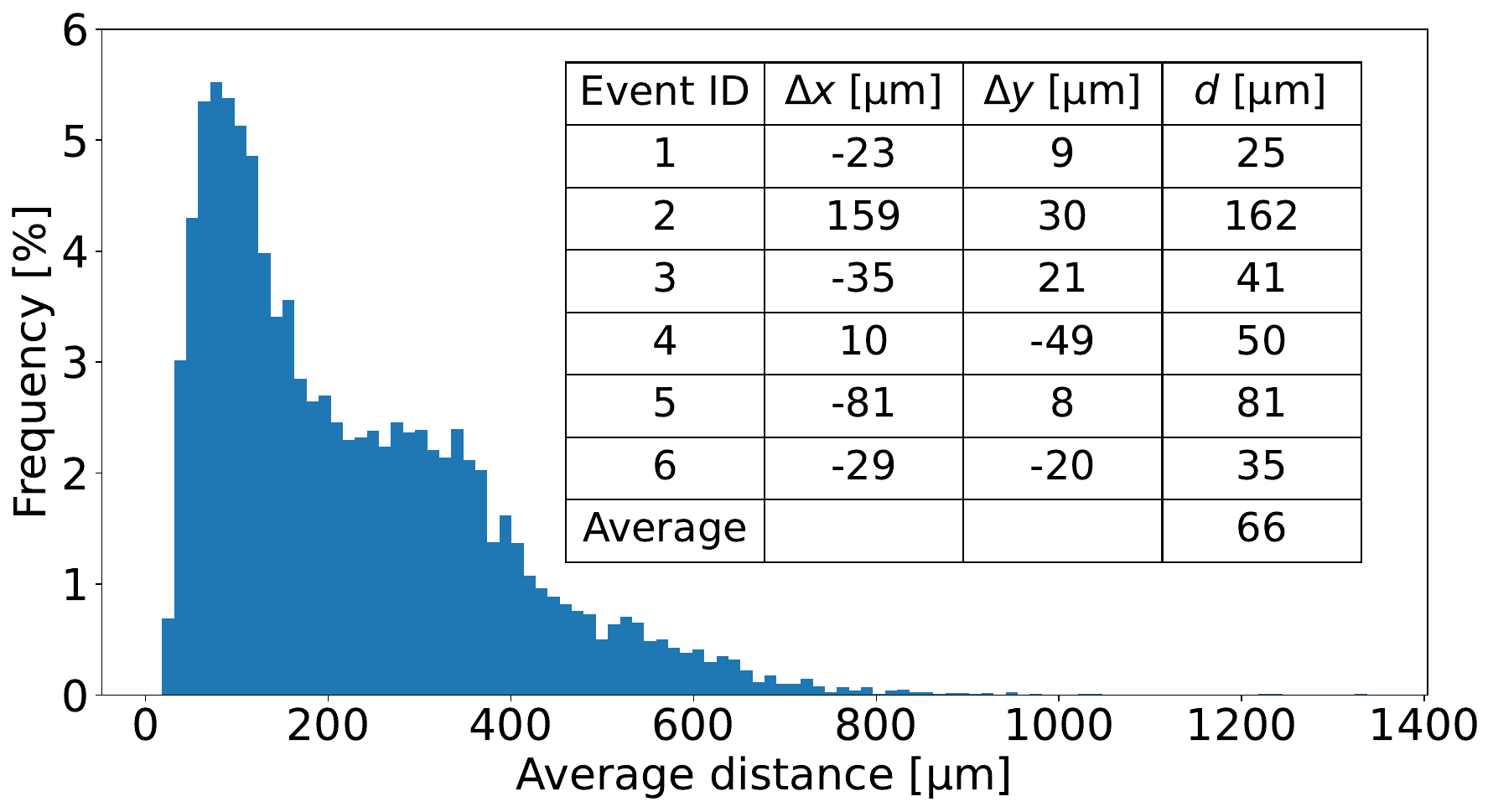}}
\caption{Histogram of the average distance between the particle hit position on the Si detector and the bit flip cell location, obtained from a PHITS-based Monte Carlo simulation. The inset table gives the measured six events, where \(\Delta x\) and \(\Delta y\) are the coordinate differences in the WL and BL directions, respectively, and \(d=\sqrt{\Delta x^2 + \Delta y^2}\) [\textmu m]. The mean of these distances is 66~\textmu m.}
\label{fig:position_sim}
\end{figure}

We listed the coordinate differences for the measured six events in the inset table of Fig.~\ref{fig:position_sim}: $ \Delta x$ and $\Delta y$ denote the differences in the WL and BL directions, respectively, and $d$ is the Euclidean distance between the bit flip cell location and the corresponding hit position on the Si detector.
The mean of these six distances is 66~\textmu m. 
To validate this result, we built a 3D model of our setup in the particle and heavy ion transport code system (PHITS) \cite{Sato02012024} better to have a reference and performed Monte Carlo simulations with $2\times10^9$ trials. 
We then sampled six proton- and secondary-ion–induced SEU events, computed their coordinate differences, and 
averaged them over \(10^4\) repetitions, with the results shown in Fig.~\ref{fig:position_sim}. 
The experimentally measured offset was found to be consistent with the simulation. 
In the histogram of Fig.~\ref{fig:position_sim}, the simulation peaks at 78~\textmu m, while the measured value is 66~\textmu m.
This small deviation of 15\% indicates that the measured value is statistically reasonable.
Table~\ref{tab:comparison} compares soft error observation systems. The most significant advancement of our system is that it achieves a Pseudo MCU probability of zero.

\begin{table}[t]
\tabcolsep = 3pt
\centering
\caption{
Comparison with Published Works.
}
\begin{tabular}{|l|c|c|c|}
\hline
 & \textbf{TNS 12’~\cite{seifert2012soft}} & \textbf{TNS 25’~\cite{10855824}} & \textbf{This Work} \\
\hline
Technology & 22nm FinFET & \makecell{12nm FinFET\\28nm Bulk} & 55nm Bulk \\
\hline
\makecell{Supply\\Voltage (V)} & 0.75 & \makecell{0.68 (12nm)\\0.75 (28nm)} & 0.9 \\
\hline
Power (W) & -- & -- & 0.77 @ 1025 MHz \\
\hline
\makecell{Memory\\(Mbit/chip)} & Tens & \makecell{28.3 (12nm)\\18.9 (28nm)} & 0.33 \\
\hline
Test Methods & Dynamic & Static & Dynamic \\
\hline
\makecell{All SRAM\\Read Time} & 10 s & Several mins & 125 ns \\
\hline
\makecell{Pseudo MCU\\Probability} & $1.6 \times 10^{-8}$\textsuperscript{*} & 
\makecell{$1.4 \times 10^{-6}$\textsuperscript{*} (12nm)\\$5.8 \times 10^{-5}$\textsuperscript{*} (28nm)} & 0 \\
\hline
\end{tabular}
\vspace{1mm}
\hspace*{0cm} 
\parbox{\linewidth}{
\textsuperscript{*}
Estimated referring to \cite{1269335} with a flux of $10^6~/\mathrm{cm}^{2}/\mathrm{s}$. 
For 12 and 22nm, $2\times10^{-10}~\mathrm{cm}^2/\mathrm{Mbit}$ was used with 15~min and 10~s exposure, respectively.  
For 28nm, $8\times10^{-9}~\mathrm{cm}^2/\mathrm{Mbit}$ and 15~min were assumed.
}
\label{tab:comparison}
\end{table}

\section{Conclusion}
This paper proposes an event-wise soft-error measurement system that integrates particle detectors with a newly developed SRAM chip capable of scanning all data in 128 cycles at a clock frequency of 540 to 1025 MHz.
An 80-MeV proton irradiation experiment  verified that the system can perform sub-ns timing syncronization and event building even while the PLL is unlocked due to radiation. Event building was achieved by determining a reset offset of -32.3556 $\mathrm{\mu}s$ to align soft error and detector hit events, while spatial synchronization was maintained within a few tens of micrometers. 

\section*{Acknowledgements}
The authors thank Mr. Takahiro Nakayama of Osaka University for his preliminary chip design and analysis.

\bibliographystyle{IEEEtran}
\bibliography{bibliography}

\end{document}